\documentclass{article} % For LaTeX2e
\usepackage{iclr2019_conference,times}

\usepackage{amsmath,amsfonts,bm}

\def\eqref#1{equation~\ref{#1}}
\def\1{\bm{1}}

\DeclareMathAlphabet{\mathsfit}{\encodingdefault}{\sfdefault}{m}{sl}
\SetMathAlphabet{\mathsfit}{bold}{\encodingdefault}{\sfdefault}{bx}{n}

\usepackage{hyperref}
\usepackage{url}
\usepackage{subfig}
\usepackage{graphicx}
\usepackage{multirow}
\usepackage{framed}
\usepackage{booktabs}
\usepackage{url}

\usepackage[justification=centering]{caption}

\title{Working women and caste in India: A study of social disadvantage using feature attribution}

\author{Kuhu Joshi \\
International Food Policy Research Institute\\
South Asia Region, New Delhi, India\\
\texttt{kuhu.joshi@cgiar.org} \\
\And
Chaitanya K. Joshi \\
School of Computer Science and Engineering \\
Nanyang Technological University, Singapore \\
\texttt{chaitanya.joshi@ntu.edu.sg} \\
}

\iclrfinalcopy

\begin{document}

\maketitle

\begin{abstract}
Women belonging to the socially disadvantaged caste-groups in India have historically been engaged in labour-intensive, blue-collar work. 
We study whether there has been any change in the ability to predict a woman's work-status and work-type based on her caste by interpreting machine learning models using feature attribution.
We find that caste is now a less important determinant of work for the younger generation of women compared to the older generation. 
Moreover, younger women from disadvantaged castes are now more likely to be working in white-collar jobs.\footnote{
Code available at \small{\url{https://github.com/chaitjo/working-women}}}
\end{abstract}

\section{Introduction}
\label{intro}

Working outside the house has historically been considered a social-stigma or a low-status activity for women in India \citep{eswaran2013status}. 
As a result, only the poorest women are working out of necessity in blue-collar jobs and once their family income increases, they withdraw from the workforce \citep{rao2010gender}. Women tend to re-enter the work force only at high education levels that allow access to un-stigmatized white-collar jobs \citep{klasen2012push}. In this cultural context, caste is an important determinant of a woman's work-status. Since scheduled castes and scheduled tribes (Sc/St) are the socially and economically disadvantaged caste groups in India, Sc/St women have always had higher workforce participation, mainly in blue-collar jobs. Upper-caste (or general caste) women have historically been discouraged from working in order to maintain a higher social-status \citep{eswaran2013status}.\footnote{
\cite{bayly2001caste} provides a comprehensive discussion of the caste system in India.}

We study whether the ability to infer a woman's work-status based on her caste is changing over generations. We also study the change in the effect of caste on work-status over generations. Adopting strategy similar to \cite{bertrand2018coming}, we use a nationally representative dataset from the National Family Health Survey \citep{nfhs4} for training ensemble Gradient Boosting Decision Tree models to predict women's work-status. Our models uncover non-linear temporal patterns between caste and women's work-status using the SHAP (SHapley Additive exPlanation) feature attribution framework \citep{lundberg2017unified}.

SHAP values of a feature (such as belonging to the Sc/St caste) measure how important that feature is in predicting the outcome of a model (woman's work-status). 
Using individual-level SHAP values, we find an upward trend across generations of working women, where caste is a less important determinant of work-status for younger women. 
Further, we unpack the impact of caste on blue-collar and white-collar jobs, finding that younger Sc/St women are moving out of blue-collar and into white-collar jobs in India.

Our paper is related to recent studies on machine learning for development \citep{de2017discovery,de2018machine,mullainathan2017machine}. We demonstrate a novel usage of the SHAP framework to study social disadvantage and how it affects women's work in India. We also contribute to the aforementioned literature on caste and work by studying for the first time whether the younger generation of women are still facing the same level of disadvantage as older women.

\section{Data}
\label{data}

We use a nationally representative dataset from the National Family Health Survey conducted in 2015-16 (NFHS-4) consisting of 699,686 women aged 15-49 years across all 29 states and 7 union territories of India \citep{nfhs4}. The survey has information on caste, work-status, and occupation types for a representative sub-sample of 111,398 women. We consider women over 21 years of age, leaving us with 81,816 women for whom we have detailed socio-economic information for our analysis. 

We classify working women into those who have blue-collar type jobs (agriculture, skilled and unskilled manual labour, and domestic services) and those who have white-collar type jobs (professional, technical, managerial, clerical, and sales). We use sixteen socio-economic features for training machine learning models to predict work-status. Appendix \ref{datastats} presents summary statistics of the dataset and the features used.

\section{Experiments}
\label{experiments}

We design three binary classification experiments to predict a woman's work: (1) having a job or not (\textit{work-status}), (2) having a blue collar job or not (\textit{blue-collar}), and (3) having a white collar job or not (\textit{white-collar}).
Using the sampling weights provided in NHFS-4, we create a test set with 5\% of the data and use the remaining 95\% for training our models in each experiment. We use stratified sampling to ensure that positive-to-negative class balance remains the same across training and test sets.

For each experiment, we train an ensemble Gradient Boosting Decision Tree (GBDT) model using LightGBM \citep{ke2017lightgbm}. 
We chose LightGBM due to generally strong empirical performance, fast training time, and easy compatibility with the SHAP framework. Optimal model hyperparameters are found using 5-fold cross-validated grid search over the training set. We use F1-score as the metric for choosing the best hyperparameters since it incorporates class imbalance. These hyperparameters are used to re-train each of the models on the entire training set. 

Next, we interpret the trained models using the SHAP feature attribution framework for tree ensembles. Unlike traditional feature importance methods, SHAP allows for fast computation of complex tree-based, non-linear models.
We compute SHAP values for our entire dataset, following \cite{lundberg2018consistent} as a guide to using and interpreting the explanations obtained. 
These explanations allow us to understand how a single feature affects the model's output, summarize relative feature importance over the entire dataset, and analyze higher order interactions among feature pairs.
% \footnote{See \url{towardsdatascience.com/interpretable-machine-learning-with-xgboost-9ec80d148d27} for an accessible overview of SHAP.}

Since we use a nationally representative dataset, results from our experiments represent actual country-level patterns. However, it is important to note that the accuracy of feature attribution methods is proportional to model performance. See Appendix \ref{performance} for a detailed discussion.

\section{Results}
\label{results}

%%%

\begin{figure}[h!]
\centering
\begin{minipage}{.5\textwidth}
\centering
  \includegraphics[width=0.85\linewidth]{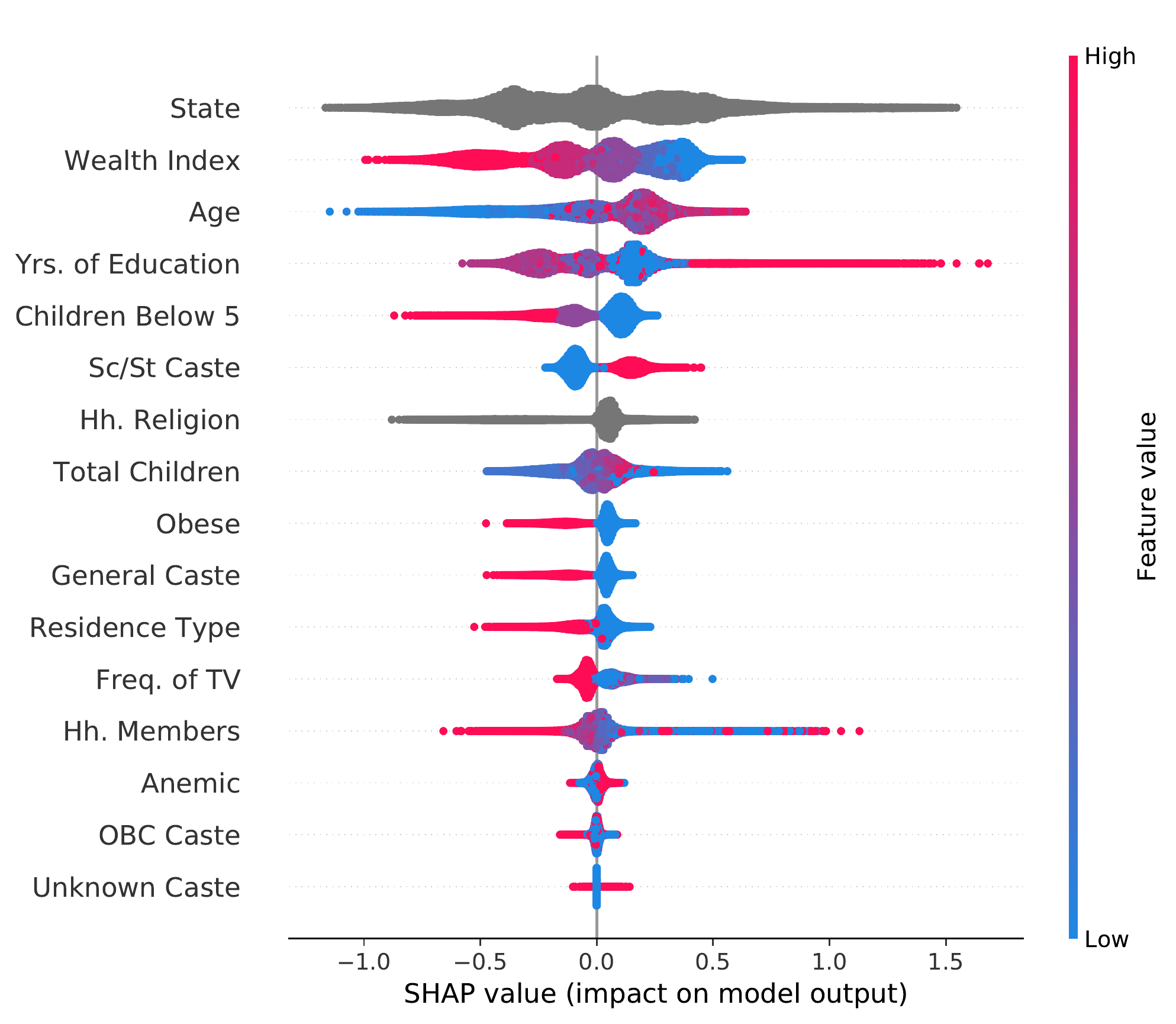}
  \caption{SHAP summary plot for \textit{work-status} experiment. Categorical variable values are grey.}
  \label{fig:summary_work}
\end{minipage}%
\begin{minipage}{.5\textwidth}
\centering
  \vspace*{3.6mm}
  \includegraphics[width=0.95\linewidth]{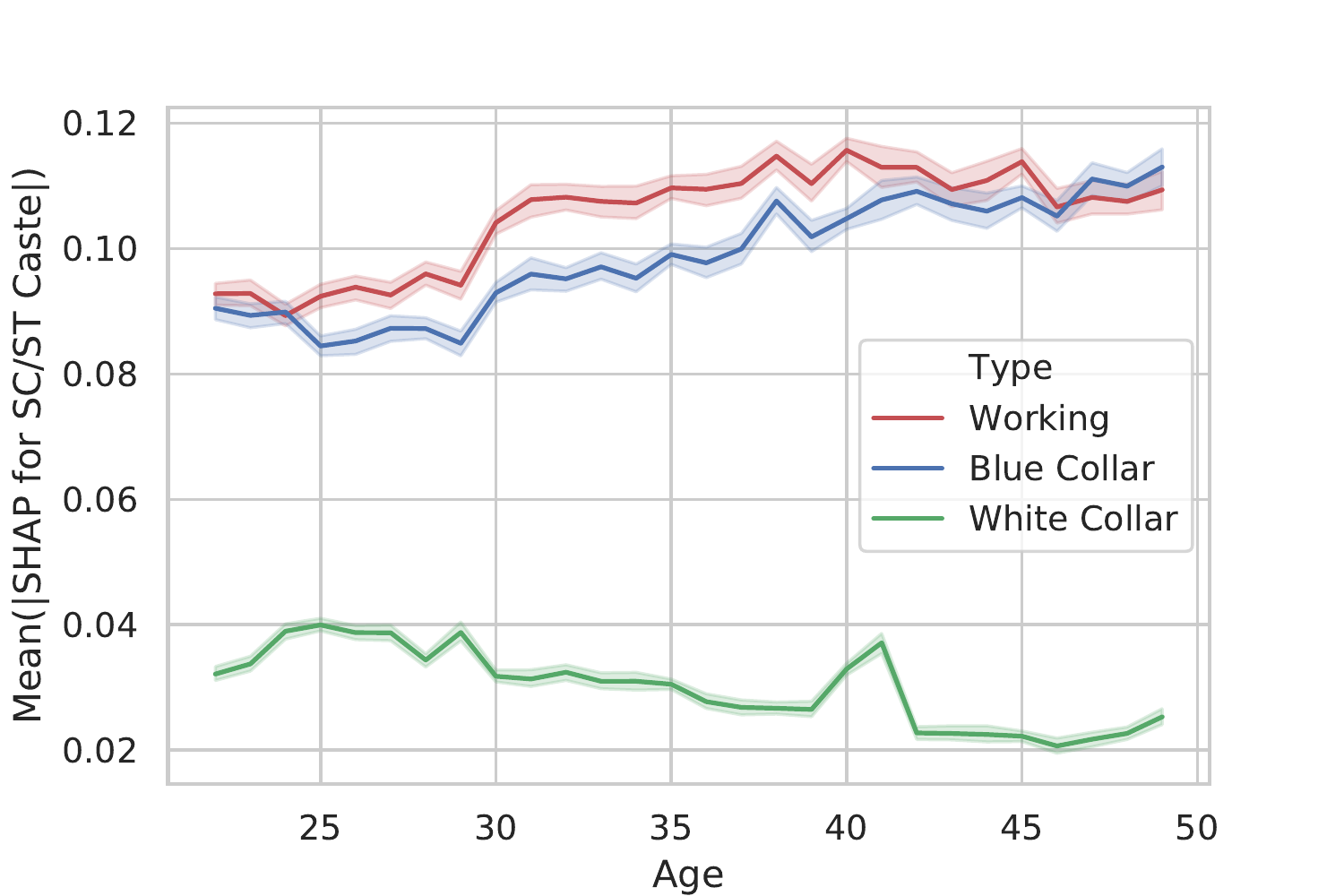}
  \vspace*{3.6mm}
  \caption{Mean of the magnitude of SHAP \\ values of Sc/St Caste over generations.}
  \label{fig:generations}
\end{minipage}
\end{figure}

\paragraph{Global Feature Importance} 
The SHAP summary plot for \textit{work-status} experiment in Figure \ref{fig:summary_work} shows the relative importance of features, the distribution of impacts of features on the model's prediction, as well as how the feature's value (Low to High) relates to its impact.
Each dot represents a woman in the dataset and the x-axis position of the dot is the impact of that feature on the model's prediction. Dots that do not fit on the row pile up to show density. Since the GBDT model has a logistic loss, the x-axis has units of log-odds.
Summary plots for \textit{blue-collar} and \textit{white-collar} experiments can be found in Appendix \ref{additional}.

Belonging to the Sc/St caste is the sixth most important feature in predicting work-status. State, household wealth index, age, years of education, and number of children below 5 years of age, are the five most important predictors of work-status. The observed patterns are similar to those found in existing literature: women from wealthier households are less likely to work \citep{rao2010gender}, more educated women are more likely to work \citep{bhalla2011labour}, and Sc/St women are more likely to work \citep{srivastava2010women}.

%%%

\paragraph{Importance of caste over generations} 
For each of the three experiments, we plot the mean of the magnitude of SHAP values of the Sc/St Caste feature for discrete ages,
along with the 99\% confidence interval,
to analyze whether the importance of caste has changed over generations (Figure \ref{fig:generations}). 
We find that caste is more important in predicting work-status of older women than of younger women. This pattern is also observed for blue-collar jobs. For white collar jobs, there isn't a clear monotonic pattern, but overall, caste is more important for women younger than 35 years of age.

%%%

\begin{figure}[h!]
\begin{minipage}{.5\textwidth}
\begin{center}
  \includegraphics[width=0.85\linewidth]{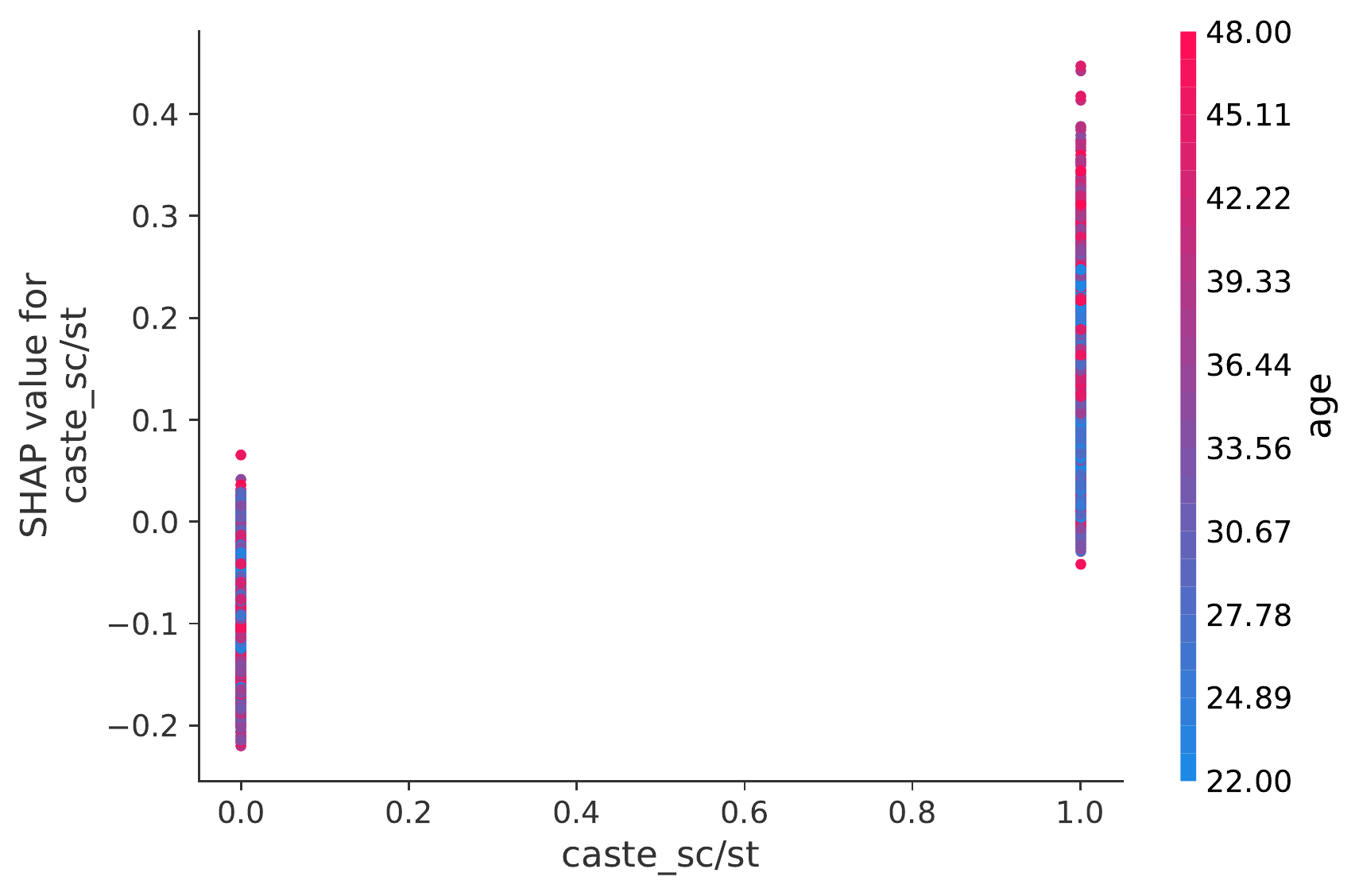}
  \caption{SHAP dependence plot of Sc/St Caste coloured by age for \textit{work-status} experiment.}
  \label{fig:dep_caste-age_work}
  \end{center}
\end{minipage}%
\begin{minipage}{.5\textwidth}
\begin{center}
  \vspace*{1mm}
  \includegraphics[width=0.7\linewidth]{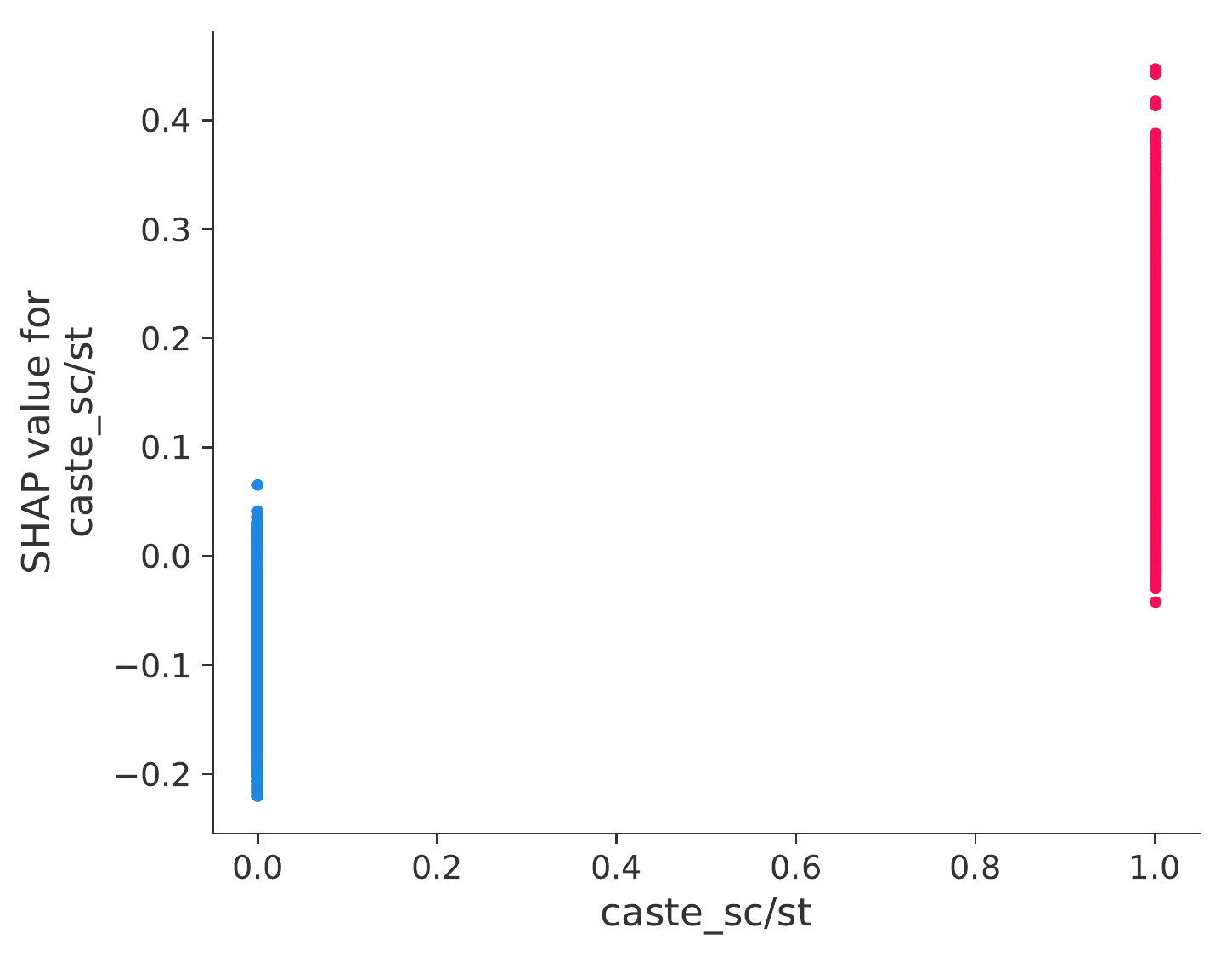}
  \vspace*{0.5mm}
  \caption{SHAP main effects of Sc/St Caste \\ for \textit{work-status} experiment.}
  \label{fig:main_caste_work}
  \end{center}
\end{minipage}
\end{figure}

\paragraph{Dependence and main effects of caste} 
The SHAP dependence plot in Figure \ref{fig:dep_caste-age_work} plots the SHAP values of the Sc/St Caste feature from the \textit{work-status} experiment and colours them by age to visualize underlying interaction effects. 
We can see the attributed importance of caste change as its value varies. Higher SHAP values represent higher probability of working: Sc/St women are more likely to be working than women of other castes. 
The dependence plot also captures vertical dispersion at a single value of the Sc/St Caste feature due to interaction effects with other features in the model. Colouring each dot by age (as the interacting feature) we find that older Sc/St women are most likely to be working.

Next, we analyze the main effects plot for the Sc/St Caste feature by removing all the interaction effects with other features in the \textit{work-status} experiment (Figure \ref{fig:main_caste_work}). We find that Sc/St women are highly likely to be working while women of other castes are not likely to be working.

%%%

\begin{figure}[h!]
\centering
\begin{minipage}{.5\textwidth}
  \centering
  \vspace*{1mm}
  \includegraphics[width=0.85\linewidth]{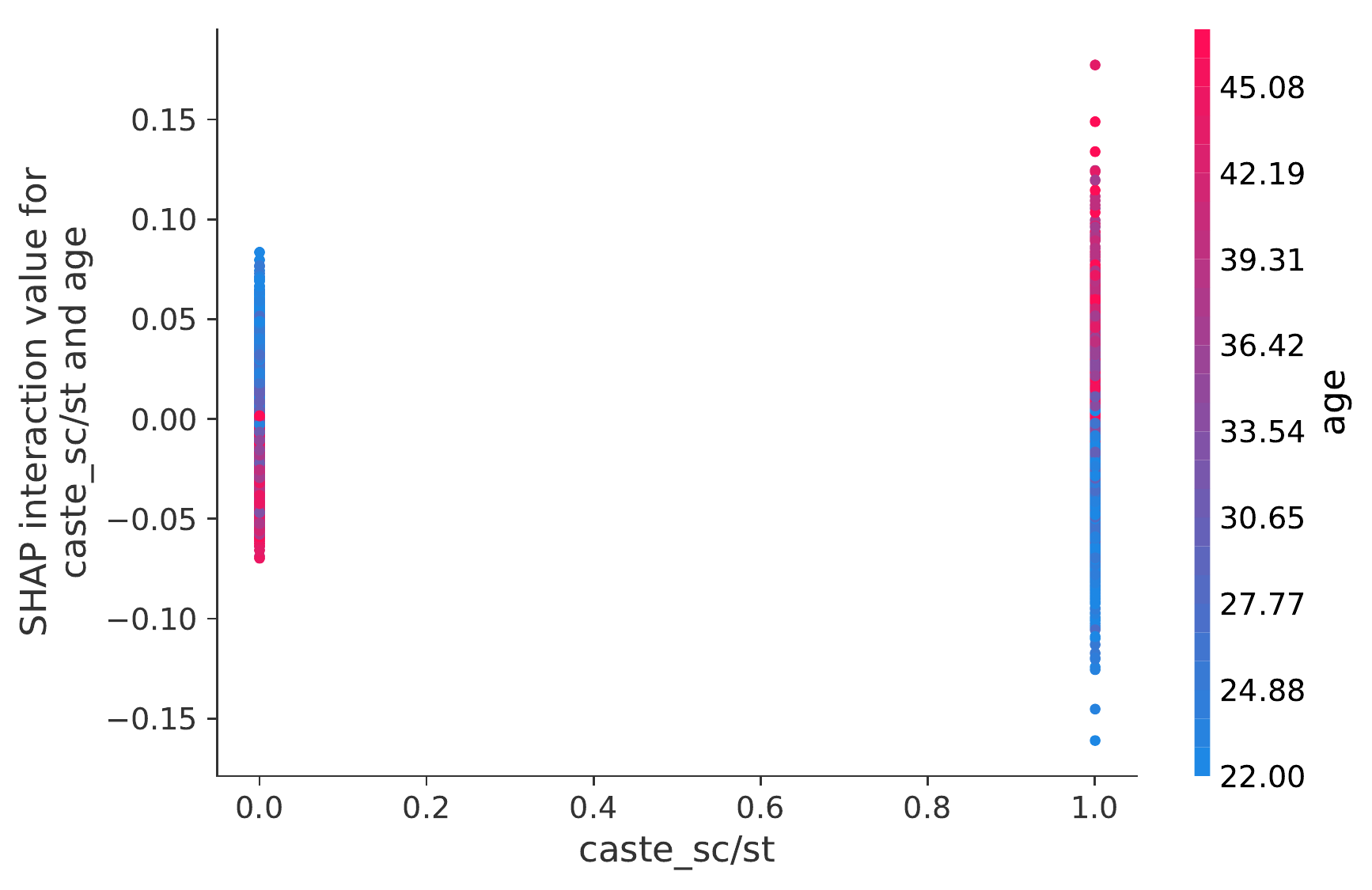}
  \caption{SHAP interaction plot of Sc/St \\ Caste and age for \textit{work-status} experiment.}
  \label{fig:int_caste-age_work}
\end{minipage}%
\begin{minipage}{.5\textwidth}
  \centering
  \includegraphics[width=0.85\linewidth]{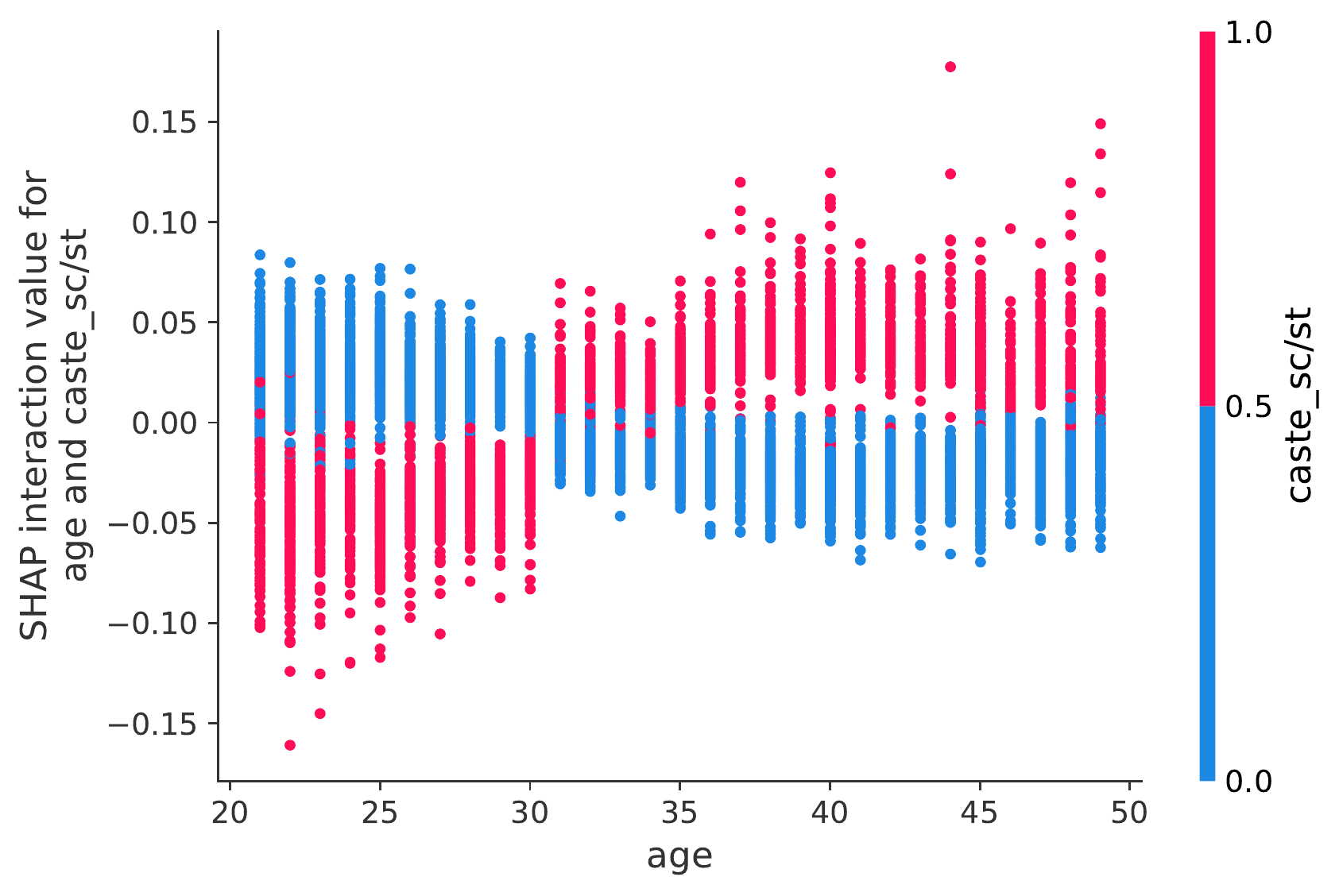}
  \caption{SHAP interaction plot of age and \\ Sc/St Caste for \textit{work-status} experiment.}
  \label{fig:int_age-caste_work}
\end{minipage}
\end{figure}

\paragraph{Interactions between caste and age}
SHAP interaction plots capture all of the vertical dispersion that was present in the SHAP dependence plot (Figure \ref{fig:dep_caste-age_work}) but was missing from the main effects plot (Figure \ref{fig:main_caste_work}). We only focus on interaction plots between the Sc/St Caste feature and age (Figure \ref{fig:int_caste-age_work}). For younger Sc/St women, SHAP interaction values are larger negatives, implying a high prediction for not-working. For older Sc/St women, SHAP interaction values are larger positives, implying a high prediction for working. Plotting the SHAP interaction values using age on the x-axis instead of caste, we get a sharper visualization of the same pattern (Figure  \ref{fig:int_age-caste_work}).

Note that SHAP interaction values are expensive to compute. Hence, main effect and interaction plots use a random sub-sample of 10,000 women sampled using the weights provided in NFHS-4.

%%%

\begin{figure}[h!]
\begin{minipage}{.5\textwidth}
\begin{center}
  \vspace*{1mm}
  \includegraphics[width=0.85\linewidth]{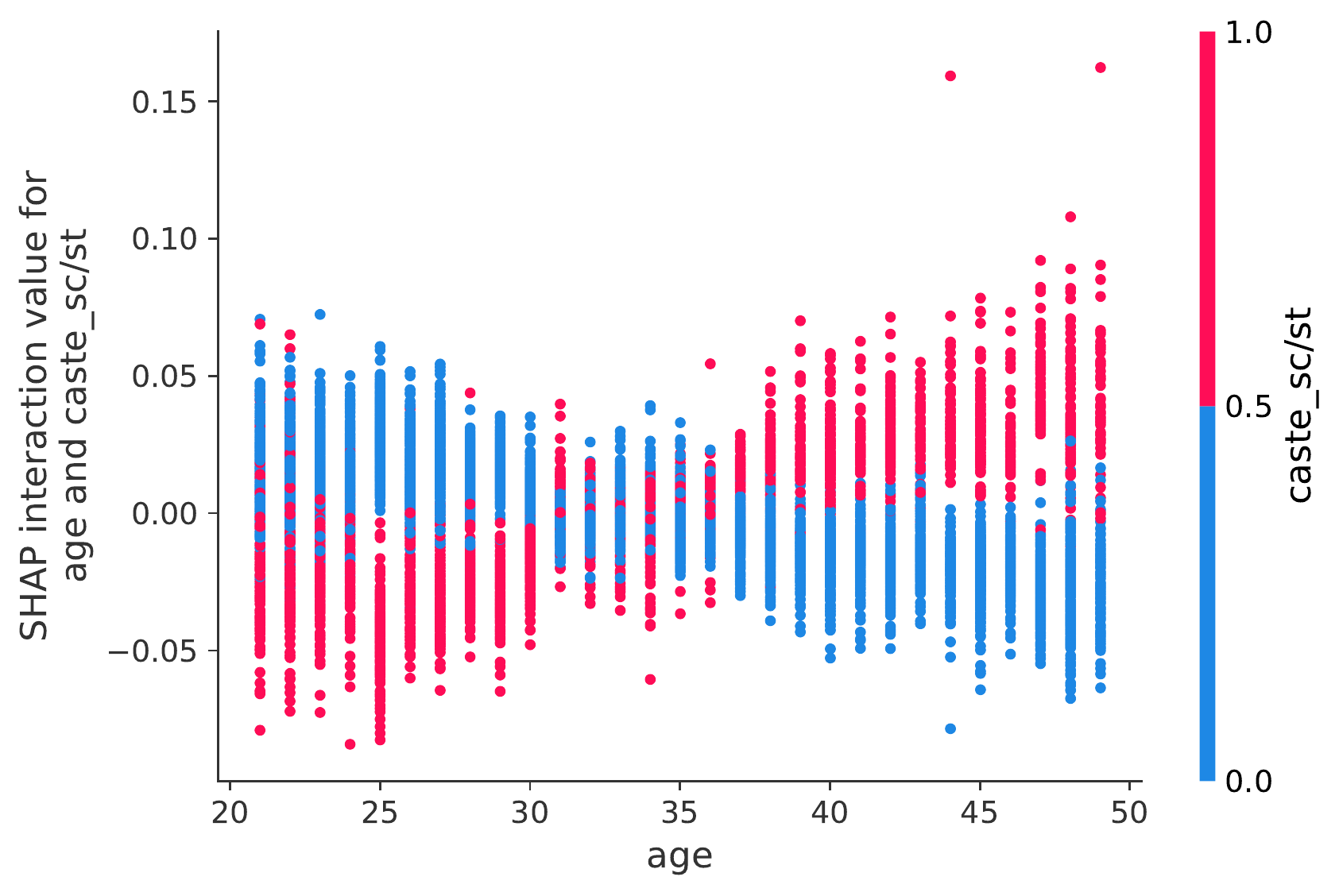}
  \caption{SHAP interaction plot of age and \\ Sc/St Caste for \textit{blue-collar} experiment.}
  \label{fig:int_age-caste_blue}
  \end{center}
\end{minipage}%
\begin{minipage}{.5\textwidth}
\begin{center}
  \includegraphics[width=0.85\linewidth]{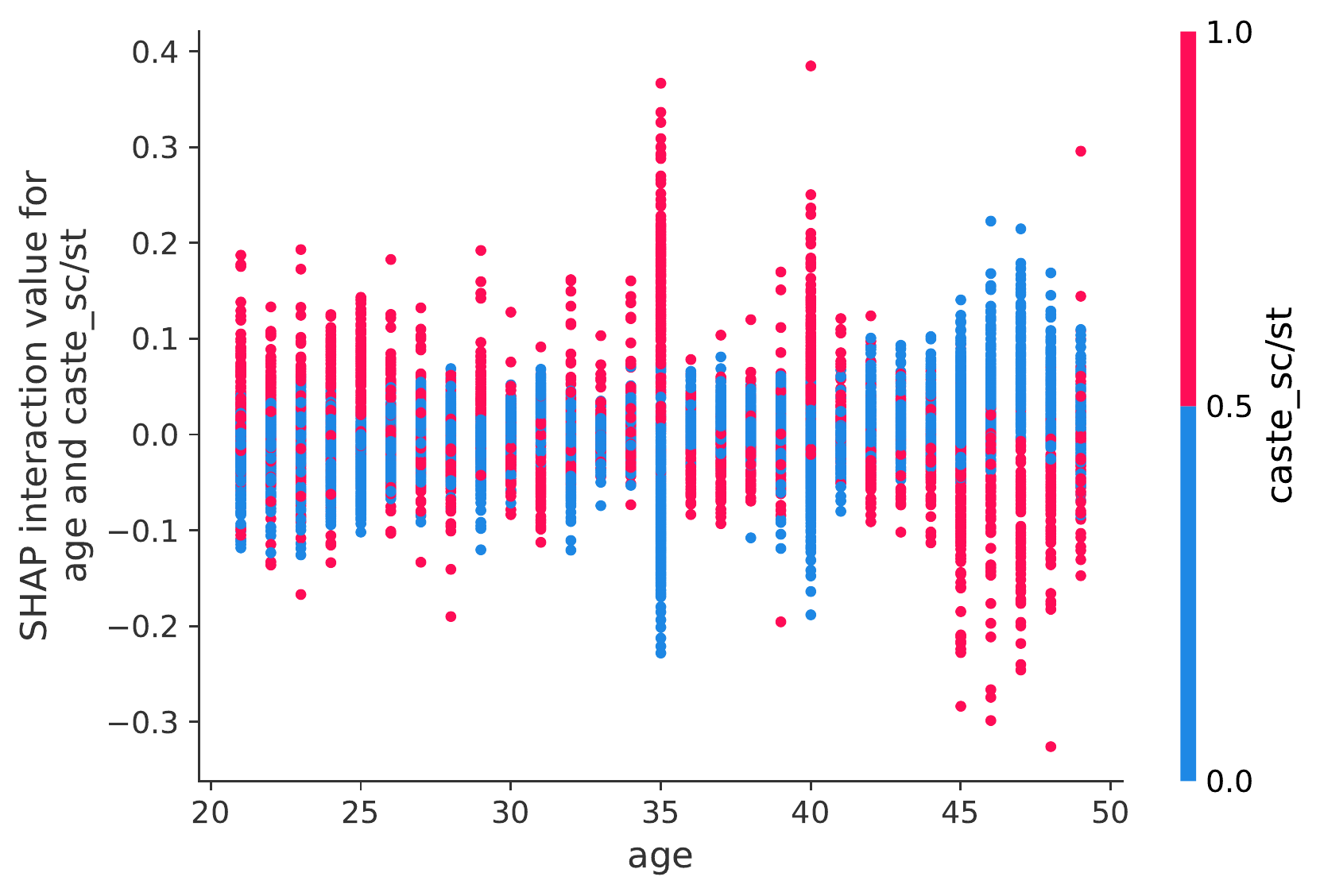}
%   \vspace*{0.5mm}
  \caption{SHAP interaction plot of age and \\ Sc/St Caste for \textit{white-collar} experiment.}
  \label{fig:int_age-caste_white}
  \end{center}
\end{minipage}
\end{figure}

\paragraph{Interactions for job types}
For fine-grained analysis, we plot the age and Sc/St Caste interaction plots for both \textit{blue-collar} and \textit{white-collar} experiments. We find that the pattern observed in the \textit{work-status} experiment is only relevant for blue-collar jobs (Figure \ref{fig:int_age-caste_blue}). We observe an opposite pattern for white-collar jobs (Figure \ref{fig:int_age-caste_white}). Younger Sc/St women are more likely to be working in white-collar jobs while older Sc/St women are less likely to be working in white-collar jobs. This pattern is prominent mainly for women younger than 26 and older than 45 years, respectively. 

\section{Conclusions}

We train machine learning models to predict women's work-status in India and analyze the relationship between work-status and caste using the SHAP feature attribution framework. 
We find that, over generations, caste has become a less important determinant of younger women's work-status, especially their participation in blue-collar jobs.
Next, we unpack the impact of caste on work-status using SHAP interaction effects, finding that younger women of scheduled castes and scheduled tribes are less likely to be working in blue collar-jobs, and more likely to be working in white-collar jobs. 

For nearly 70 years, the government of India has implemented caste-based quotas in education and government jobs to ensure equal economic opportunity. Our work can be used as a cost-effective tool to monitor the impact of the existing caste-quota policy. In future work, we will look deeper into geographical variations to identify regions where younger women are lagging, and where job creation can be targeted.
Our methodology can similarly be used to study the persistence of, as well as the nuanced patterns underlying other types of social disadvantage and bias in both developing and developed country contexts.

\newpage

\section*{Acknowledgment}
We would like to thank Maria De-Arteaga, Scott Lundberg, Sean Saito and the anonymous reviewers for helpful feedback and discussions that have been included in the Addendum (Appendix \ref{addendum}).

\small
\bibliography{iclr2019_conference}
\bibliographystyle{iclr2019_conference}

\normalsize
\newpage

\appendix
\section{Dataset Statistics}
\label{datastats}

Table \ref{table:data} shows the distribution of various castes across work-status types in the representative sample dataset of women above 21 years of age. Overall, 34.1\% of women are working. 28.3\% of women are doing blue-collar jobs and 5.8\% are employed in white-collar jobs. 42.7\% of scheduled caste or scheduled tribe women are working, while only 23.9\% of general caste women are working.

\begin{table}[h!]
\small
\centering
\caption{Distribution of number of women from each caste across work-status types.}
\label{table:data}
\begin{tabular}{lccccc}
\toprule
Work-status & General Caste & Sc/St Caste & OBC Caste & Unknown Caste &  Total \\
\midrule
Unemployed & 13,978 & 17,480 & 22,254 & 230 & 53,942 \\
Blue-collar & 3,068 & 11,235 & 8,730 & 108 & 23,141 \\
White-collar & 1,341 & 1,787 & 1,592 & 13 & 4,733 \\
\bottomrule
\end{tabular}
\end{table}

Table \ref{table:features} presents the summary statistics of the sixteen socio-economic features provided as inputs to machine learning models: each woman's personal biographical data including age, years of education, state of residence and caste (split into four binary features); information about her household such as type of residence (rural/urban), household religion, household wealth index and household members; and her health/reproductive history represented by number of children ever born, number of children currently under five years of age, whether she is anemic, whether she is obese, and her frequency of watching TV.

Among these features, we are particularly interested in looking at the relationship between work-status and caste, which may take four values: general caste, Sc/St caste, other backward class (OBC), or unknown. To facilitate detailed analysis using feature attribution, we split caste into four binary features for each woman: whether she is from the general caste, whether she is from the Sc/St caste, whether she is from the OBC caste, and whether her caste is unknown. Having a positive value for any one of these four features means that the values of the other features are zeros.

\begin{table}[h!]
\centering
\caption{Summary statistics for the features used as inputs to machine learning models.}
\label{table:features}
\resizebox{\textwidth}{!}{%
\begin{tabular}{lcccccccl}
\toprule
{Feature} & {Age} & {Yrs. of Education} & {State} & {Residence Type} & {Hh. Religion} & {Wealth Index} & {Hh. Members} & {Freq. of TV} \\
\midrule
{Type} & Numeric & Numeric & Categorical & Binary & Categorical & Numeric & Numeric & Numeric \\
{Values/Range} & {[}21, 49{]} & {[}0, 20{]} & 36 Categories & \{0, 1\} & 10 Categories & {[}0, 4{]} & {[}1, 39{]} & {[}0, 3{]} \\
{Mean} & 33.765 & 6.301 & - & 0.309 & - & 2.088 & 5.546 & 2.047 \\
{Std. Dev.} & 8.153 & 5.401 & - & 0.462 & - & 1.389 & 2.553 & 1.256 \\
\bottomrule
\\
\toprule
{Feature} & {Total Children} & {Children Below 5} & {Anemic} & {Obese} & {General Caste} & {Sc/St Caste} & {OBC Caste} & \multicolumn{1}{c}{{Unknown Caste}} \\
\midrule
{Type} & Numeric & Numeric & Binary & Binary & Binary & Binary & Binary & \multicolumn{1}{c}{Binary} \\
{Values/Range} & {[}0, 15{]} & {[}0, 9{]} & \{0, 1\} & \{0, 1\} & \{0, 1\} & \{0, 1\} & \{0, 1\} & \multicolumn{1}{c}{\{0, 1\}} \\
{Mean} & 2.426 & 0.632 & 0.520 & 0.234 & 0.224 & 0.372 & 0.398 & \multicolumn{1}{c}{0.004} \\
{Std. Dev.} & 1.749 & 0.907 & 1.389 & 0.499 & 0.417 & 0.483 & 0.489 & \multicolumn{1}{c}{0.065} \\
\bottomrule
\end{tabular}%
}
\end{table}

%%%

\section{Model Performance and SHAP Explanations}
\label{performance}

Table \ref{table:performance} presents the performance of the three GBDT models described in Section \ref{experiments} on the respective training and test sets. We evaluate our models based on binary accuracy and F1-score.
Models for \textit{work-status} and \textit{blue-collar} perform acceptably and do not overfit to the training set. However, F1-score for the \textit{white-collar} model clearly indicates that it has overfit to the training set and generalizes poorly to unseen data. We believe this is due to extremely low class balance.

It is important to note that the SHAP frameworks explains how GBDT models work, but the models are not guaranteed to be causal. Outcomes of predicting work-status may have been impacted by several hidden factors not provided to the model or for which data collection was impossible. Hence, model performance and generalization to unseen data is a good indicator of how much we should trust the explanations produced by SHAP. In practice, we observe that using the sampling weights from NFHS-4 leads to explanations that are coherent with existing literature on gender and labour in India.

\begin{table}[h!]
\begin{center}
\caption{Training and test set performance for GBDT models. Class balance is computed as number of samples from the positive class divided by total number of samples.}
\label{table:performance}
\begin{tabular}{lccccc}
\toprule
\multirow{2}{*}{Experiment} & \multirow{2}{*}{Class Balance} & \multicolumn{2}{c}{Training Set} & \multicolumn{2}{c}{Test Set} \\
& & Accuracy & F1-score & Accuracy & F1-score \\ 
\midrule
\textit{work-status} & 0.341 & 0.692 & 0.604 & 0.673 & 0.578 \\
\textit{blue-collar} & 0.283 & 0.716 & 0.596 & 0.681 & 0.551 \\
\textit{white-collar} & 0.058 & 0.898 & 0.510 & 0.853 & 0.305 \\
\bottomrule
\end{tabular}
\end{center}
\end{table}

\section{Additional Visualizations}
\label{additional}

\begin{figure}[h!]
\centering
\begin{minipage}{.5\textwidth}
\centering
  \includegraphics[width=0.85\linewidth]{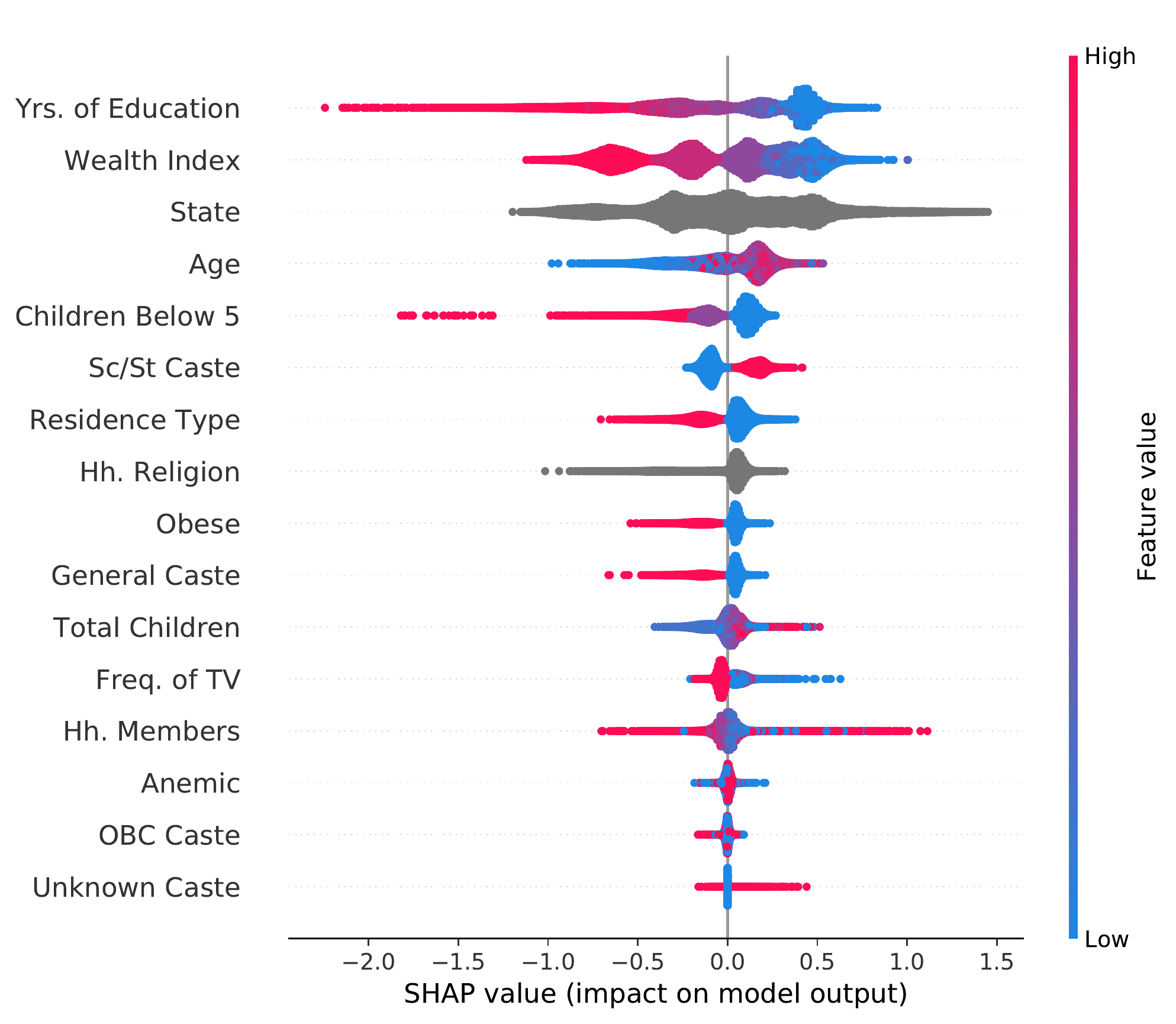}
  \caption{SHAP summary plot for \\ \textit{blue-collar} experiment.}
  \label{fig:summary_blue}
\end{minipage}%
\begin{minipage}{.5\textwidth}
\centering
  \includegraphics[width=0.85\linewidth]{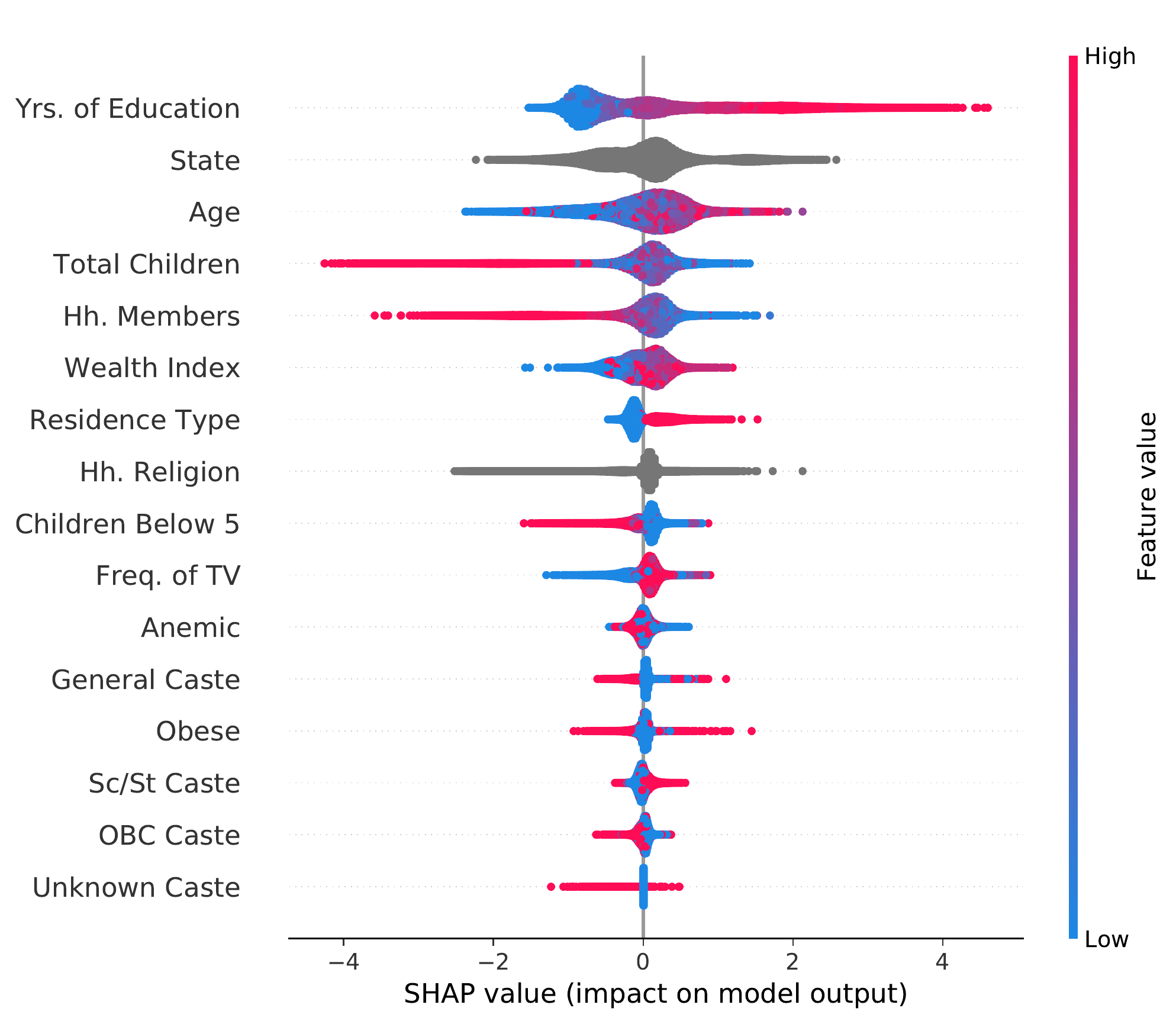}
  \caption{SHAP summary plot for \\ \textit{white-collar} experiment.}
  \label{fig:summary_white}
\end{minipage}
\end{figure}

\paragraph{Summary plots for work-types}
Figures \ref{fig:summary_blue} and \ref{fig:summary_white} show the SHAP summary plot for \textit{blue-collar} and \textit{white-collar} experiments, respectively. 
For blue-collar jobs, belonging to the Sc/St caste is the sixth most important feature after years of education, household wealth index, state, age, and number of children below 5 years of age. Highly educated women or those from wealthy households are unlikely to be employed in blue-collar jobs.

Years of education, state, age, total number of children, and household members are the five most important predictors of white-collar employment. None of the features related to caste are important predictors. Highly educated women are likely to be employed in white-collar jobs, whereas women with many children or large household sizes are not.

%%%

\begin{figure}[h!]
\centering
\begin{minipage}{.5\textwidth}
\centering
  \vspace*{9.5mm}
  \includegraphics[width=0.85\linewidth]{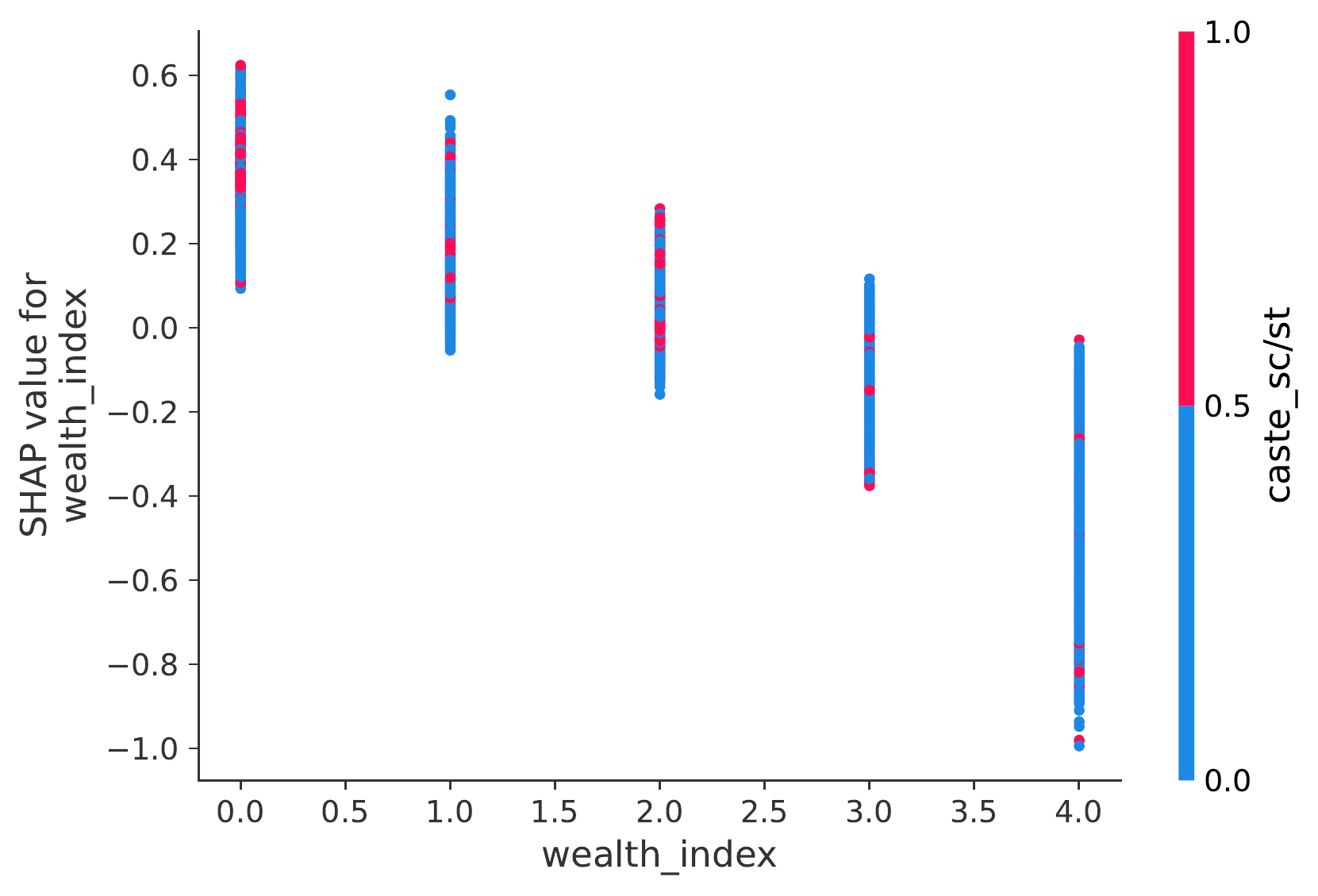}
  \vspace*{9.5mm}
  \caption{SHAP dependence plot of household wealth index coloured by Sc/St Caste for \textit{work-status} experiment.}
  \label{fig:dependence_wealth}
\end{minipage}%
\begin{minipage}{.5\textwidth}
\centering
  \includegraphics[width=0.85\linewidth]{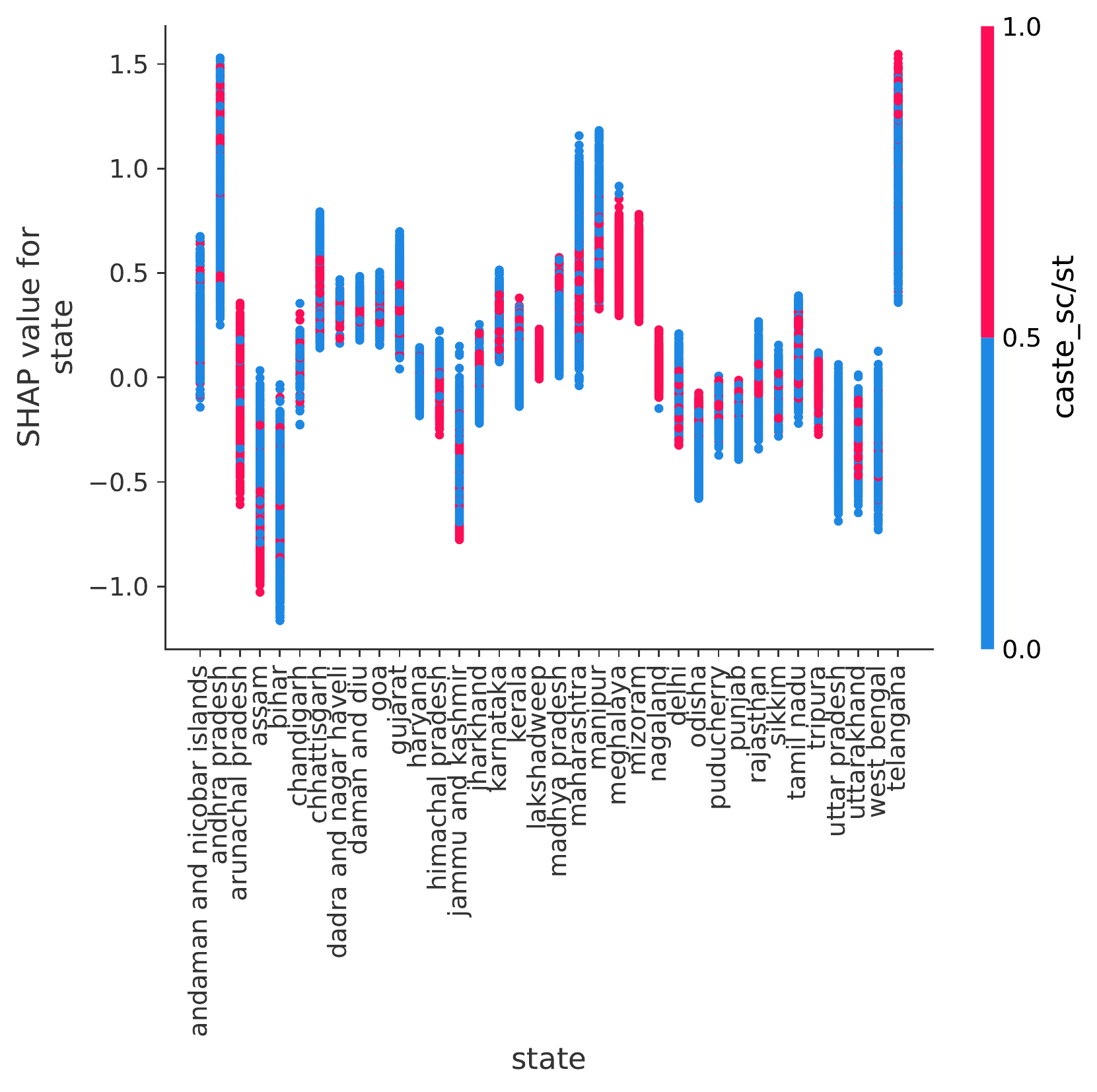}
  \caption{SHAP dependence plot of state coloured by Sc/St Caste for \textit{work-status} experiment.}
  \label{fig:dependence_state}
\end{minipage}
\end{figure}

\paragraph{Household wealth and work-status}
The dependence plot of household wealth index in Figure \ref{fig:dependence_wealth} shows that most of the Sc/St women in the dataset are from the bottom three wealth levels. For women belonging to wealthier households, our model gives a high prediction for not-working. This is consistent with socio-economic literature on women's work in India \citep{mammen2000women}.

\paragraph{State-wise patterns of work-status}
The dependence plot in Figure \ref{fig:dependence_state} shows differences in the caste-composition of working women across the states of India. It also shows state-wise differences in the likelihood of working. Women in Andhra Pradesh, Telangana, Maharashtra, Manipur, and Mizoram are most likely to be working. Women in Bihar, Assam, and Jammu and Kashmir are least likely to be working. 

\begin{figure}[h!]
\centering
    \begin{framed}
    \includegraphics[width=.85\linewidth]{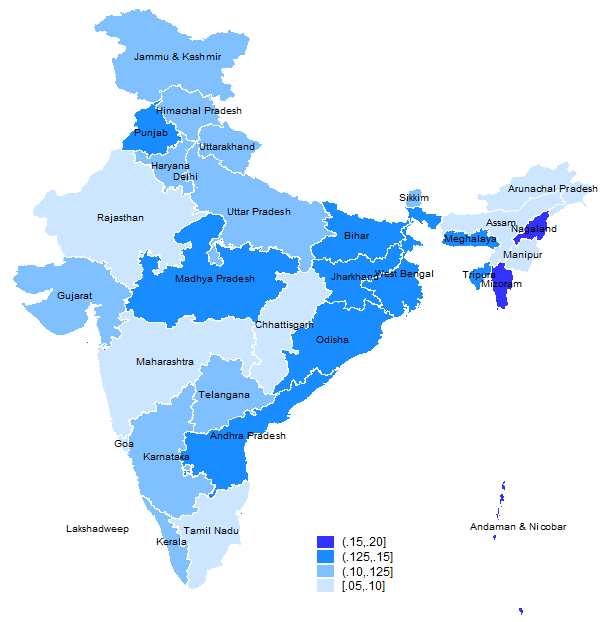}
    \end{framed}
    \caption{Mean of the magnitude of SHAP values of Sc/St Caste \\ across states of India for \textit{work-status} experiment}
    \label{fig:state_mean_SHAP}
\end{figure}

Figure \ref{fig:state_mean_SHAP} plots the mean of the magnitude of SHAP values of Sc/St Caste feature across states of India. It shows the distribution of the importance of caste in predicting work-status across states.

\newpage

\newpage

\begin{figure}[h!]
    \centering
    \includegraphics[width=0.85\linewidth]{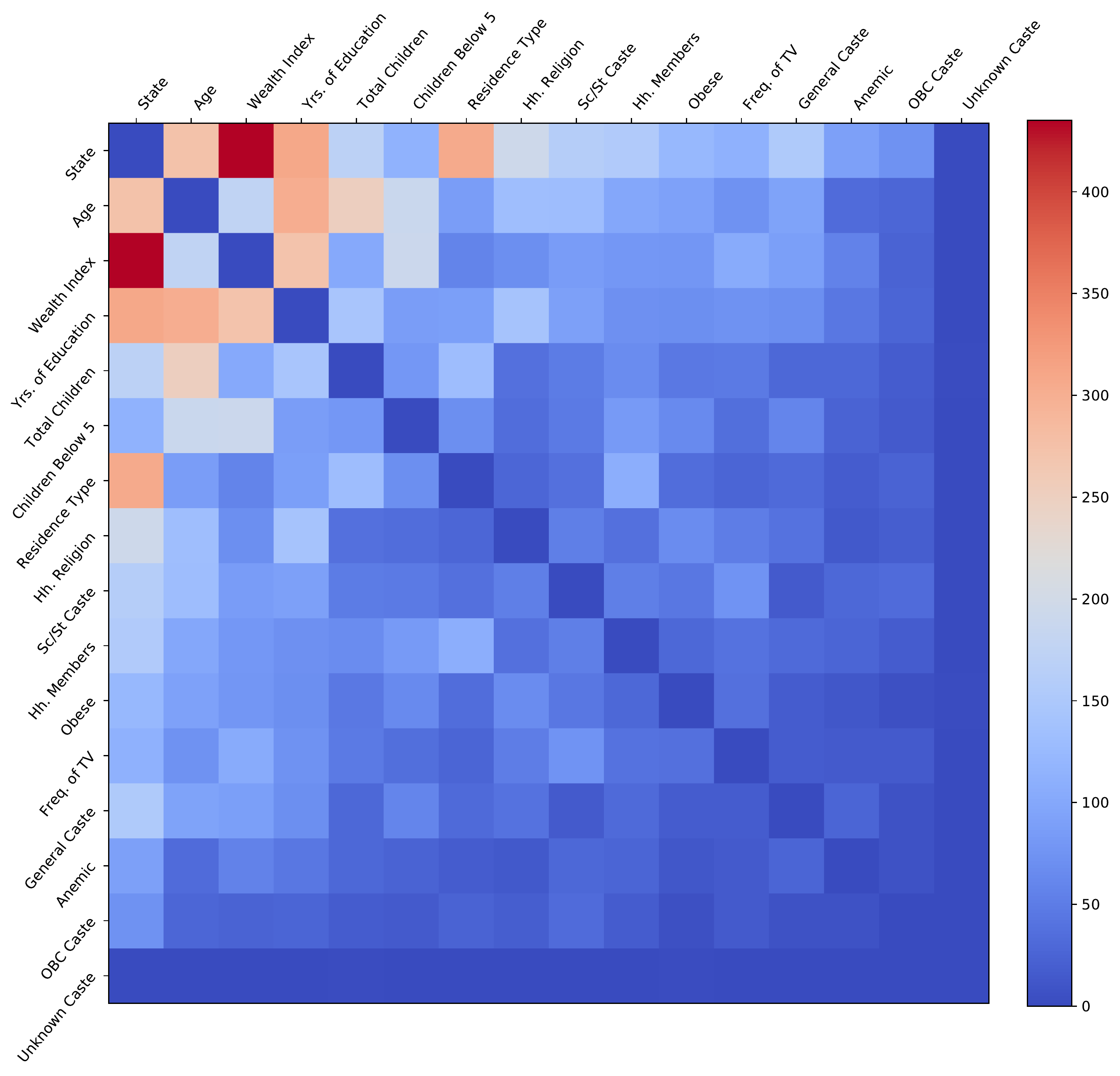}
    \caption{Heatmap of magnitude of SHAP interaction values for \textit{work-status} experiment. \\ Shades from cool to warm denote increasing strength of interaction.}
    \label{fig:heatmap_work}
\end{figure}

\paragraph{Interaction heatmap}
Figure \ref{fig:heatmap_work} shows a heatmap of magnitude of SHAP interaction values for the \textit{work-status} experiment.
SHAP interaction values are a generalization of SHAP values to higher order interactions. The SHAP framework computes a matrix for every prediction, where the main effects are on the diagonal and the interaction effects are off-diagonal. 
The main effects are similar to the SHAP values for a linear model. The interaction effects capturing all the higher-order interactions are divided up among the pairwise interaction terms. 

The heatmap shows that pairwise interactions between the more predictive features for the model, such as state, wealth index and years of education, are the most important at capturing interaction effects. For the Caste Sc/St feature, we see that state and age are the most important interaction effects.  

%%%
\section{Addendum}
\label{addendum}

Since the initial presentation of this work, we have received helpful feedback on our motivation and methodology, as well as on the literature on interpretable machine learning.
We hope the following discussions are useful for future work in the intersection of social issues, demographic datasets and model interpretation using feature attribution (\textit{e.g.} \cite{vilarino2019using}).

\paragraph{How much can we trust SHAP explanations?}
A broad motivation for this work was to use SHAP to discover non-linear patterns and relationships in nationally representative socio-economic datasets.
In Appendix \ref{performance}, we discussed how performance on the predictive task used for training our model was a good indicator of the quality of downstream explanations using applying SHAP to the model's prediction.
However, the relationship between model performance and SHAP explanations is not obvious.
Recent work \citep{lipton2016mythos,rudin2019stop,lakkaraju2019fool,slack2019can} has highlighted several shortcomings of post-hoc explanation methods such as SHAP, 
leading us to be more cautious about our methodology and conclusions.

In our experiments, models trained on a subset of the full dataset were used to obtain explanations for the full dataset.
Although we used a nationally representative dataset and found explanations to be coherent with existing literature, 
a large difference in size between the training and test  sets (95\% and 5\%, respectively) could result in conclusions that are modelling the noise in the training data.
Thus, we propose a simple robustness check to ascertain that the patterns hold for unseen data and are not `overfit' to the training set:
Given a large dataset, we can bootstrap several random subsets of the dataset for training models. 
We can then compare the coherence of the explanations for samples not used in training with the explanations for the full dataset.

\begin{figure}[h!]
\centering
\begin{minipage}{.5\textwidth}
\centering
  \vspace*{8mm}
  \includegraphics[width=0.95\linewidth]{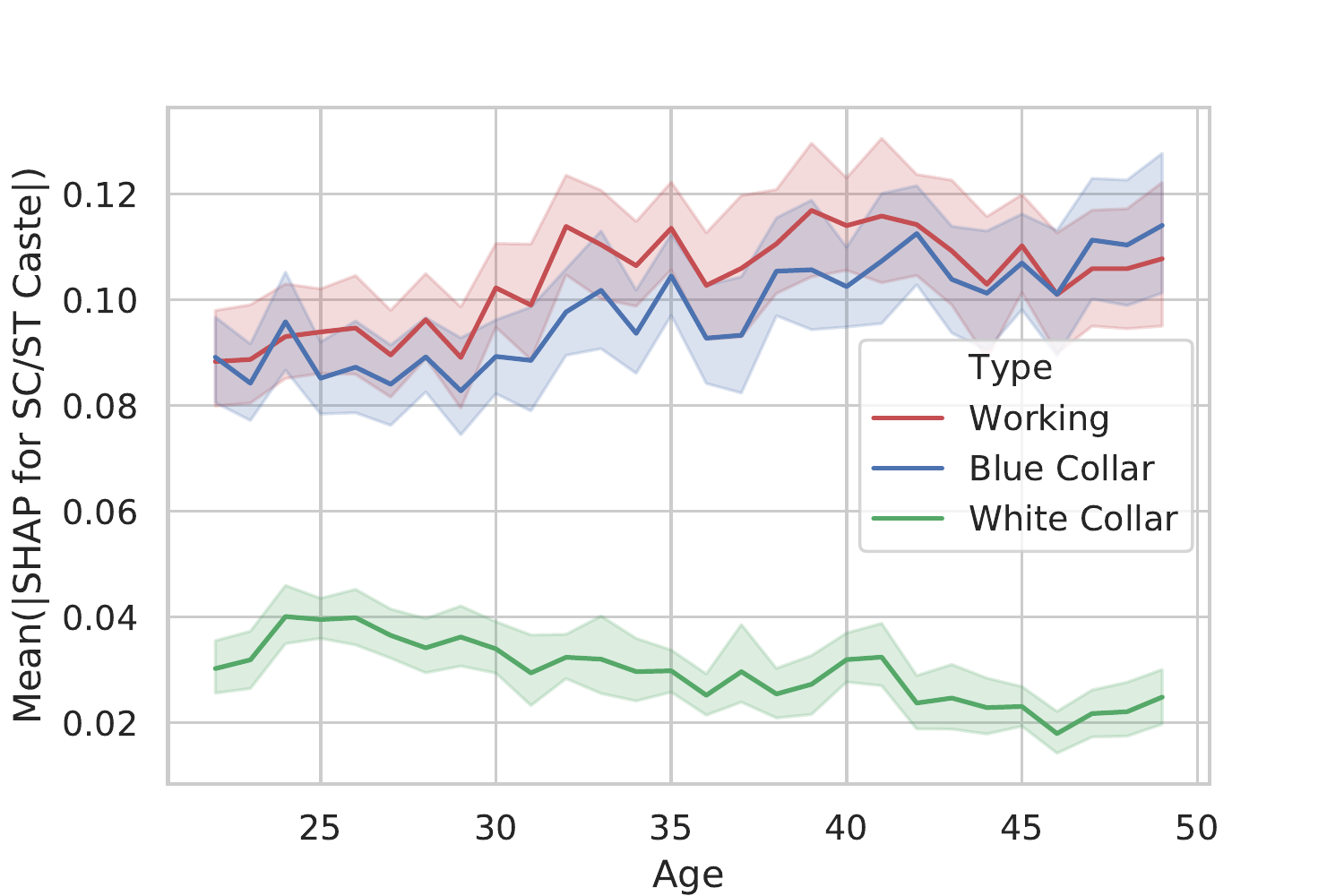}
  \vspace*{9.25mm}
  \caption{Mean of the magnitude of SHAP \\ values of Sc/St Caste over generations \\ (Test set only).}
  \label{fig:generations-test}
\end{minipage}%
\begin{minipage}{.5\textwidth}
\centering
  \includegraphics[width=0.95\linewidth]{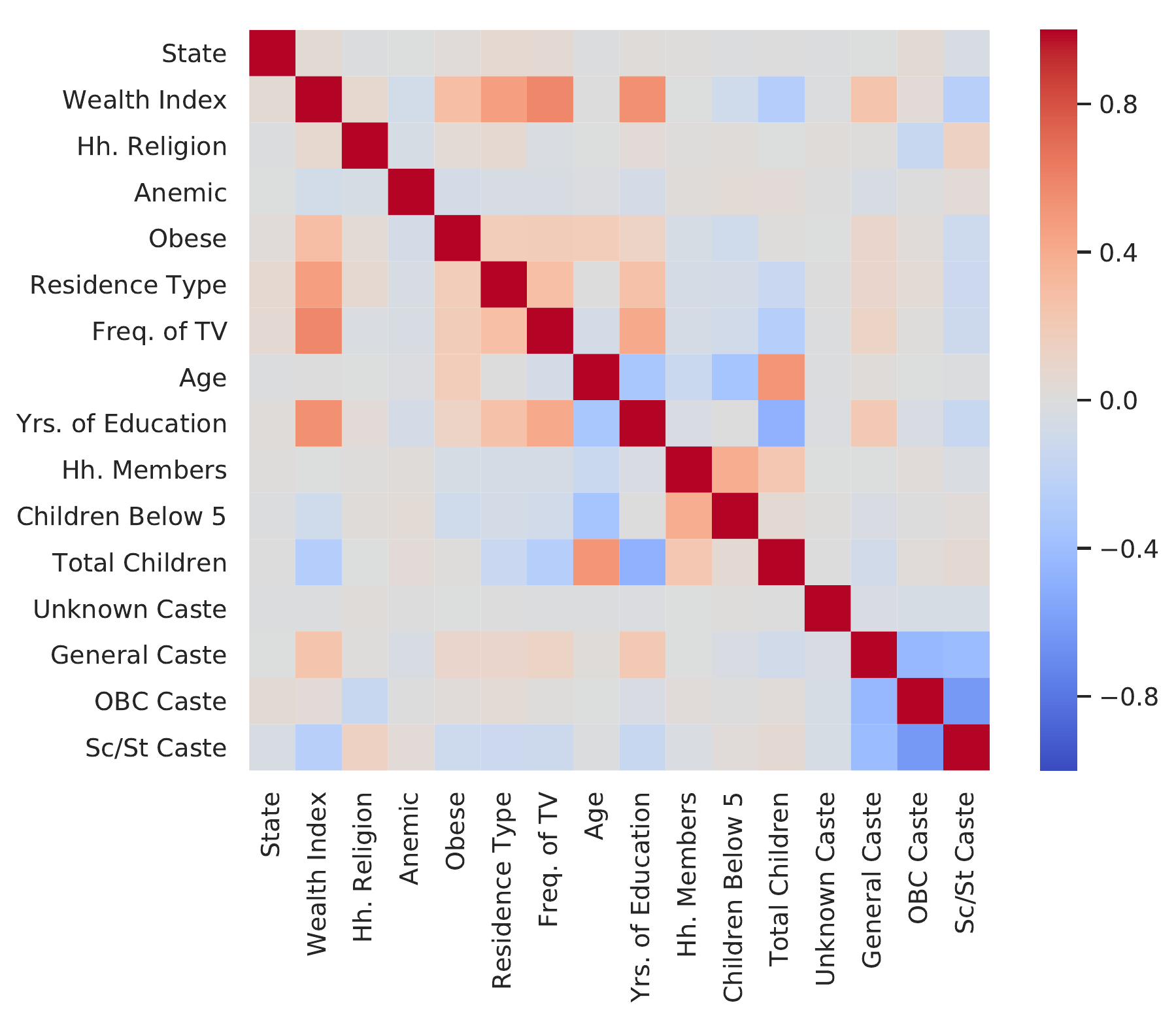}
  \caption{Spearman's rank correlation heatmap for socio-economic features from NFHS-4.}
  \label{fig:spearman}
\end{minipage}
\end{figure}

For example, Figure \ref{fig:generations-test} shows the mean of the magnitude of SHAP values for the Sc/St Caste feature across ages for the held-out test set only.
Comparing Figure \ref{fig:generations-test} with Figure \ref{fig:generations}, we observe similar patterns for the importance of caste over generations, albeit with higher variance in SHAP values for the test set.

\paragraph{Dealing with correlated features}
Selecting input features for training models, which are later explained using SHAP, poses an interesting challenge:
if two input features are closely correlated, the model might consistently use the more predictive feature while ignoring the other. 
Hence, explanations based on feature attribution might show the other feature having no predictive power, which is not necessarily true.

In Figure \ref{fig:spearman}, we use Spearman's rank correlation\footnote{
Spearman's rank correlation measures how well the relationship between two variables can be described using a monotonic function.}
as a measure of correlation for our dataset, and found features such as the household wealth index and years of education, or the number of children and age, to be highly correlated.
Although no features were highly correlated with caste, we found clear trends in caste's correlation with household wealth, years of education and type of residence.
General caste shows small positive correlation with higher levels of wealth and education as well as urban residences, whereas the opposite trend is seen for Sc/St caste.
Thus, we believe that choice of input features and their correlations are an important consideration when analyzing model explanations from SHAP.

\end{document}